\documentclass[twocolumn]{aastex62}
\usepackage[T1]{fontenc}
\usepackage{amsmath,amssymb}
\usepackage{epstopdf}
\usepackage{ulem}
\usepackage{multirow}
\usepackage{todonotes}
\usepackage{lineno}

\usepackage{soul}
\usepackage{float}
\usepackage{xcolor}
\usepackage[utf8]{inputenc} 

\newcommand{\mpo}{\textcolor{black}}

\newcommand{\pjm}{\textcolor{black}}
\newcommand{\pjmii}{\textcolor{black}}

\shorttitle{Electron-Whistler Interactions in Oblique Shocks}
\shortauthors{Morris et al.}

\def\ee{\end{equation}}
\def\be{\begin{equation}}

\newcommand{\thbn}{\theta_\mathrm{Bn}}

\newcommand{\ompe}{\omega_\mathrm{pe}}

\newcommand{\omce}{\Omega_\mathrm{ce}}
\newcommand{\omci}{\Omega_\mathrm{ci}}

\newcommand{\mi}{m_\mathrm{i}}
\newcommand{\me}{m_\mathrm{e}}
\newcommand{\lse}{\lambda_\mathrm{se}}
\newcommand{\lsi}{\lambda_\mathrm{si}}
\newcommand{\wce}{\Omega_\mathrm{ce}}
\newcommand{\vsh}{v_\mathrm{sh}}
\newcommand{\vpar}{v_{\parallel}}

\newcommand{\vperp}{v_{\bot}}
\newcommand{\veth}{v_{e,\rm{th}}}
\newcommand{\ppar}{p_{\parallel}}
\newcommand{\pperp}{p_{\bot}}
\newcommand{\xsh}{x_{\rm sh}}
\newcommand{\kpar}{k_{\parallel}}
\newcommand{\kperp}{k_{\bot}}
\newcommand{\rpathp}{R_{\rm{path}}'}
\newcommand{\rlinp}{R_{\rm{lin}}'}
\newcommand{\roscp}{R_{\rm{osc}}'}
\newcommand{\tlinp}{t_{\rm{lin}}'}

\newcommand{\omres}{\varpi_{\mathrm{res}}}

\newcommand{\ti}{T_\mathrm{i}}
\newcommand{\te}{T_\mathrm{e}}

\defcitealias{Morris2022}{Paper I}

\begin{document}

\title{Pre-acceleration in the Electron Foreshock II: Oblique Whistler Waves}

\correspondingauthor{Paul J. Morris}
\email{paul.morris@desy.de}

\author[0000-0002-8533-8232]{Paul J. Morris}
\affil{Deutsches Elektronen-Synchrotron DESY, Platanenallee 6, 15738 Zeuthen, Germany}

\author[0000-0002-5680-0766]{Artem Bohdan}
\affil{Deutsches Elektronen-Synchrotron DESY, Platanenallee 6, 15738 Zeuthen, Germany}
\affiliation{Max-Planck-Institut für Plasmaphysik, Boltzmannstr. 2, DE-85748 Garching, Germany}

\author[0000-0002-3440-3225]{Martin~S.~Weidl}
\affiliation{Max-Planck-Institut für Plasmaphysik, Boltzmannstr. 2, DE-85748 Garching, Germany}

\author[0000-0003-3417-1425]{Michelle Tsirou}
\affil{Deutsches Elektronen-Synchrotron DESY, Platanenallee 6, 15738 Zeuthen, Germany}

\author[0000-0001-6002-6091]{Karol Fulat}
\affil{Institute of Physics and Astronomy, University of Potsdam, D-14476 Potsdam, Germany}

\author[0000-0001-7861-1707]{Martin Pohl}
\affil{Deutsches Elektronen-Synchrotron DESY, Platanenallee 6, 15738 Zeuthen, Germany}
\affil{Institute of Physics and Astronomy, University of Potsdam, D-14476 Potsdam, Germany}

\begin{abstract}

Thermal electrons have gyroradii many orders of magnitude smaller than the finite width of a shock, thus need to be pre-accelerated before they can cross it and be accelerated by diffusive shock acceleration. One region where pre-acceleration may occur is the inner foreshock, which upstream electrons must pass through before any potential downstream crossing. In this paper, we perform a large scale particle-in-cell simulation that generates a single shock with parameters motivated from supernova remnants. Within the foreshock, reflected electrons excite the oblique whistler instability and produce electromagnetic whistler waves, which co-move with the upstream flow and as non-linear structures eventually reach radii of up to 5 ion-gyroradii. We show that the inner electromagnetic configuration of the whistlers evolves into complex non-linear structures bound by a strong magnetic field around 4 times the upstream value. Although these non-linear structures do not in general interact with co-spatial upstream electrons, they resonate with electrons that have been reflected at the shock. We show that they can scatter, or even trap, reflected electrons, confining around $0.8\%$ of the total upstream electron population to the region close to the shock where they can undergo substantial pre-acceleration. This acceleration process is similar to, yet approximately 3 times more efficient than, stochastic shock drift acceleration.

\end{abstract}


\section{Introduction}\label{introduction}

It was suggested by \citet{Fermi1949} that hadrons could be accelerated by magnetic mirrors and give rise to the well-documented observed cosmic-ray power-law spectrum \citep[e.g.][]{1984ARA&A..22..425H,Nagano_2009}. The original Fermi acceleration assumes interactions between particles and magnetic mirrors, which move at speed $U$, occur with an isotropic distribution of incident angles between the two (as measured by a stationary observer). \pjm{The relative orientation of these interactions can either cause particles to gain (when head-on) or lose (if head-tail) energy, with a slight preference (of order $U/c$) for the former. The expected angle-averaged fractional energy gain per collision is} $\propto (U/c)^2$, where $c$ is the speed of light. Astrophysical shocks occurring in nature are more accurately described by diffusive shock acceleration (DSA) \citep{1977DoSSR.234.1306K,1977ICRC...11..132A,Bell1978a,1978ApJ...221L..29B}, where the interactions always occur head-on in the rest-frames both upstream (unshocked plasma) and downstream (shocked plasma) of the shock front. DSA analytically predicts a more efficient \pjm{fractional} energy-gain-per-crossing of $\propto (U/c)$ and power law spectrum. DSA has been generally successful in explaining a wide variety of astrophysical sources such as active galactic nuclei \citep{Marchenko2017} and supernova remnants (SNRs) \citep{2008ARA&A..46...89R}, where we observe non-thermal emission that is often characterized by a power-law. The radiative properties of these objects can be interpreted as originating from an underlying population of high-energy particles (protons, electrons etc.), providing evidence to support DSA.

Despite its numerous successes, \pjm{aspects of} the underlying micro-physics necessary for DSA to work are \pjm{yet to be conclusively determined} \citep{2022RvMPP...6...29A}. \pjm{This is because in DSA it is assumed} that the shock is a perfect discontinuity, when in reality it has a finite width of the order of \pjm{the gyroradius of a proton traveling with the shock speed}, $r_{gi}$. While this does not pose challenging for thermal ions to cross into the downstream, thermal electrons require a significant amount of pre-acceleration before their gyroradii \pjm{are sufficiently enlarged} that they can \pjm{easily} cross the shock transition \pjm{from upstream to} downstream, \pjm{or vice-versa. To accelerate particles to cosmic-ray energies, DSA requires them to cross the shock multiple times, thus only electrons that have already undergone sufficient pre-acceleration can ``be injected" into DSA and undergo further acceleration by this mechanism.} 

Particle-in-cell (PIC) simulations are an excellent tool with an eminent track record when it comes to investigating electron pre-acceleration. These simulations are fully kinetic, containing individual electrons and ions, thus allow for a self-consistent treatment when these particles move in their self-generated electromagnetic fields. From these, we can obtain time- and spatially dependent information concerning both individual particles and the fields they experience which allow us to unveil the underlying physical processes \citep{Pohl2020}. 

In this work, we use physical parameters appropriate for supernova remnants, which are characterized by non-relativistic outflows with sonic and Alfv\'enic Mach numbers of $M_S, M_A \approx 20 -2000$ \citep{2009ApJ...699L.139W}. In contrast, the low Mach number regime is associated with the Earth's bow shock ($M_s, M_A < 10$). Focusing on SNR parameters is advantageous for many reasons. First and foremost, it has been established for almost 50 years that cosmic rays (CRs) can be accelerated by SNRs, with the majority of Galactic CRs believed to originate from these objects \citep{1977ICRC...11..132A,1977DoSSR.234.1306K,1983RPPh...46..973D,Bell1978a,1978ApJ...221L..29B}. Additionally the close proximity of SNRs permits the study of non-thermal radiation in radio, X-, and $\gamma$-rays. This radiation is often attributed to a population of accelerated electrons, thus understanding their acceleration is essential to comprehend the radiative properties of SNRs.

A crucial parameter in governing the behavior of a shock is the obliquity angle, $\thbn$, which subtends the upstream magnetic field with the shock normal. Typically perpendicular shocks, where $\thbn = 90^{\circ}$, have well defined shock transitions, with a small (of order $\sim r_{gi}$) shock foot region leading up to the ramp. These shocks have been thoroughly studied over the last decade using 2D PIC simulations \citep{2009ApJ...690..244A,Kato2010,Matsumoto2012,Matsumoto2013,Matsumoto2015,Wieland2016,Bohdan2017,Bohdan2019a,Bohdan2019b,Bohdan2020a,Bohdan2020b,Bohdan2021}. Conversely, decreasing the shock obliquity angle more freely permits the escape of energetic particles back upstream as their trajectories are tied to the magnetic field lines, allowing them to outrun the shock if sufficient energization has taken place. The extended regions containing the reflected particles are known as the electron and ion foreshocks (depending on the particle species) \citep[e.g.][]{Burgess1995,Fitzenreiter1995,Treumann2009} where the energy transported upstream by these reflected particles can excite instabilities and generate turbulence which can in turn influence, and possibly pre-accelerate upstream electrons. All upstream electrons that eventually encounter the shock must first pass through the foreshock, thus a physical description of these regions is essential to fully comprehend the overall description of electron pre-acceleration.


Prior work has demonstrated that energetic electrons that have been pre-energized by shock surfing acceleration are more likely to be reflected back upstream \citep{Amano2007}, where mirror reflection (also called shock drift acceleration, SDA) \citep{1984JGR....89.8857W,1984AnGeo...2..449L} is the mechanism responsible for the reflection \citep{2005MNRAS.362..833H}. \pjm{This latter mechanism operates on electrons gyrating close enough to the shock ramp so that part of their gyrational orbit enclose the region with enhanced magnetic field, causing a temporary orbital tightening and causing them to drift along the shock} \citep{1984JGR....89.8857W,1984AnGeo...2..449L}. Results obtained from using 1-dimensional PIC simulations demonstrated that the energy content of these reflected electrons was sufficient to power electrostatic and electromagnetic waves in the shock foot, which effectively trap electrons allowing them to undergo more cycles of shock drift acceleration (SDA), \pjm{gaining more} energy and \pjm{eventually cross into the} downstream \pjm{region} \citep{Xu2020,2021ApJ...921L..14K}.  

\citet{Bohdan2022} used 2D3V (2 spatial dimensions and all three velocity and field components) PIC simulations, with a combination of large scale and periodic boundary condition simulations, accompanied by analytically solving the dispersion to elucidate the exact instabilities excited by reflected electrons in the electron foreshock. The first of these are electrostatic electron acoustic waves (EAW). In paper I of this series, \citet[][hereafter \citetalias{Morris2022}]{Morris2022} investigated the effect of changing the orientation of the upstream magnetic field on the foreshock structure by performing a series of narrow box simulations. It was found that EAWs are quickly excited within a few ion gyro-radii, and the EAWs are stronger for decreasing $\thbn$. It was shown that these waves can interact with, and in $\sim 1\%$ of cases, divert upstream electrons away from the shock. \citet{Bohdan2022} further identified electromagnetic whistler waves in the inner foreshock region, which require a comparatively larger energy density of reflected electrons relative to EAWs, subsequently excited at later times than EAWs. These whistler waves occur on spatial scales approaching ion length scales, as opposed to the EAWs where the characteristic size is similar to the electron inertial length. In this paper we focus on the micro-physics of individual electrons which encounter these whistler waves, and interpret their behavior in the context of electron pre-acceleration in astrophysical shocks. In Section \ref{sec:setup} we outline our simulation setup before providing an overview of the shock structure in Section \ref{sec:shockStruc}. Section \ref{sec:Whistler} contains the main discussion, where we outline the properties and development of whistler waves before detailing their evolution and interaction with electrons present in the foreshock. 



\section{Simulation Setup} \label{sec:setup}

In \citetalias{Morris2022}, we investigated the effect of changing the obliquity angle, $\theta_{Bn}$ and the plane-angle, $\phi$, which characterize the orientation of the initial large-scale upstream magnetic field, on the electron foreshock at short times. The run-time of these simulations can be quantified in terms of the ion-gyrofrequency, defined as $\omci = |e|B_0/\mi$, for electron charge \pjmii{magnitude} $|e|$, magnetic field amplitude $B_0$ and ion mass $\mi$, with the total run-time $t_{\rm sim} = 7.8 \omci^{-1}$. They further employed a narrow box, spanning $4.8 \lsi$ transversely, where $\lsi$ is the ion skin length. This relatively small transverse size reduced the computational expense of a single simulation, and therefore permitted multiple simulations to be performed. The chosen parameters also additionally allowed for a comparison to the 3D simulations of \citet{Matsumoto2017}. 

Conversely, in this paper, we investigate electron pre-acceleration as a consequence of the electromagnetic foreshock, which begins to emerge at around $t_{\rm sim} \sim 10 \omci^{-1}$. This is characterized by the presence of whistler waves, which at late times develop into non-linear structures that can reach up to approximately $1-10 \lsi$ in diameter, which are better captured by our 2D3V simulations in the out-of-plane ($\phi=90^{\circ}$) case. Accordingly, to adequately resolve these structures we perform a simulation with a wider box, of transverse size $41.6 \lsi$ and with a longer total run-time of $t_{\rm sim} = 51.5 \omci^{-1}$, allowing us to follow their long-term evolution. This computationally expensive simulation featured in \citet{Bohdan2022}, and the setup will be briefly outlined below. 

Our code is a modified version of TRISTAN \citep{Buneman1993}, which cyclically solves Maxwell's equations for fields defined on a simulation grid and updates the positions of individual particles located in the grid cells according to Lorentz forces via the Vay solver \citep{Vay2008}. This version allows us to track the progression and properties of individual particles and measure the local field strengths they encounter to better elucidate the physical processes they experience. 

At the beginning of the simulation, we initialize a plasma slab by injecting ions and electrons, where $\mi$ and $\me$ are the ion and electron mass \mpo{and $\mi/\me = 50$}, co-spatially into the simulation box with the number of particles per cell per species given as $n_{0} = 40$. The two particle species are initialized in thermal equilibrium, such that $k_B T_e = k_B T_i = 9.86\cdot 10^{-4}~\me c^2$, where $k_B$ is the Boltzmann constant and $c$ the speed of light. This defines the sound speed as $c_{\rm s}=\sqrt{2 \Gamma k_BT_{\rm i}/\mi} = 0.0081c$, for adiabatic index $\Gamma = 5/3$. Across this plasma slab, we apply a large-scale, uniform, magnetic field according to $\vec B_0 = B_0 (\cos \thbn, \sin \thbn \cos \varphi,\sin  \thbn \sin \varphi) = B_0 (0.5, 0, \sqrt{3}/2)$, where $\thbn = 60^{\circ}$ and $\phi=90^{\circ}$. 

Such a setup defines the important temporal scales in our simulation, such as the electron plasma and gyrofrequencies, $\ompe$ and $\omce$ respectively. Their ratio is quantified by $\omce = |e|B_0/\me = 0.06~\ompe$, where $\ompe$ is the electron plasma frequency. From this, we can define the electron skin length, $\lse=c/\ompe=8\Delta$, that is resolved by 8 grid cells (denoted by $\Delta$). For ions, their inertial length scales as $\lsi = \sqrt{\mi/\me}~ \lse$. We ensure that these relevant frequencies are sufficiently resolved in our simulation by advancing it in time-step units of $\delta t=1/16\,\ompe^{-1}$.

We move our plasma slab with a bulk velocity as measured in the simulation frame of $\vec v_{\mathrm{up}}/c = - 0.20 \hat x$. Consequentially, a motional electric field is produced and defined by $\vec E_0 = - \vec v_{\rm{up}} \times \vec B_0$. Such a setup leads to a large value of $\nabla \times \vec{E}$, and a correspondingly large $\partial \vec{B} / \partial t$, at $x=0$ which can induce a large initial transient. We mitigate this by tapering the \mpo{initial} upstream field values to zero over the region $x \leq 50 \Delta$ \citep{Wieland2016}. \mpo{The nonzero value of $\nabla \times \vec{B}$ is} exactly compensated by a drift current carried by the ions, which is removed at $x=0$.

\begin{figure}
    \centering
    \includegraphics[width=\columnwidth]{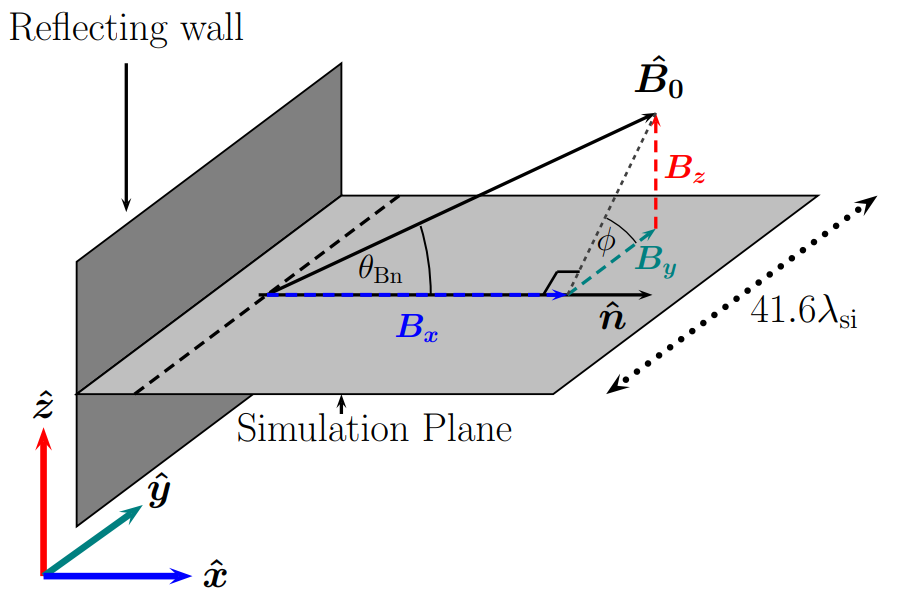}
    \caption{Reflecting wall setup of the simulation. $\phi$ is the angle $\boldsymbol{{B_0}}$ makes relative to the simulation plane. $\thbn$ is the angle subtended between the shock normal, $\boldsymbol{\hat{n}}$, and the magnetic field vector, $\boldsymbol{{B_0}}$. Here we use $\thbn=60^{\circ}$ and $\phi=90^{\circ}$, which define the upstream magnetic field as \mbox{$\boldsymbol{B_0} =  B_0( 0.5, 0, \sqrt{3}/2)$}. The dotted line indicates the transverse size of the simulation box, which is 41.6 ion skin lengths.}
    \label{fig:simSetup}
\end{figure}

Constituent particles within the plasma that reach the boundary at $x=0$ encounter a reflecting wall, which performs the transformation $v_x \rightarrow -v_x$ ($v_y$ and $v_z$ are unaffected) \citep{Quest1985,Burgess1989}, and so they propagate back upstream. The magnetic field behind this upstream-moving plasma completely isotropizes after approximately a few ion gyro times, with a compression ratio of $n_{\mathrm{d}}/n_0 \sim 4.0$ relative to the undisturbed upstream plasma. In the simulation frame, a quasi-stationary shock with velocity ${\vec v}_{\rm sh}^\ast/c = 0.067~\hat x$ propagates upstream, which is equivalent to a shock velocity of ${\vec v}_{\rm sh}/c = 0.263$ when measured in the upstream rest frame. The properties of the shock can be further quantified by the Alfv\'en velocity, $v_{\rm A}=B_{\rm 0}/\sqrt{\mu_{\rm 0} (n_e\me+n_i\mi)}$, for vacuum permeability $\mu_{\rm 0}$ and $n_i=n_e=n_0$ are the ion and the electron number densities. This gives rise to the Alfv{\'e}nic Mach number, $M_A = v_{\rm sh}/v_A = 30$, whereas the sonic Mach number is $M_S = v_{\rm sh}/c_s = 32.5$. The plasma beta value, which denotes the thermal-to-magnetic energy density ratio in the upstream region is $\beta = 1$. Our simulation setup is shown in Fig. \ref{fig:simSetup}. As the shock propagates upstream, the domain length in the \mpo{$x$-direction} increases, with new plasma injected into the new regions with the same properties outlined above. The extension of the simulation box is essential to ensure that that we maintain all reflected electrons within the boundaries of our simulation.

\section{Shock Structure and Summary of Foreshock Characteristics} \label{sec:shockStruc}

\begin{figure*}
    \centering
    \includegraphics[width=\textwidth]{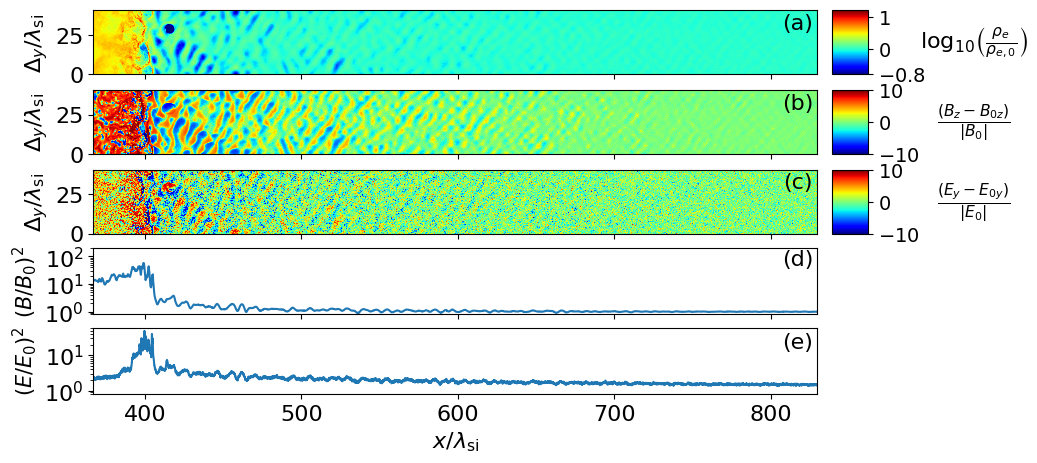}
    \caption{
     Late time shock structure for $\omci t = 49.9$.  Panel (a) shows the electron density map, (b) and (c) show fluctuations in the $B_z$ and $E_y$ components of the magnetic- and electric-fields, respectively, and (d) and (e) the magnetic and electric field profiles, respectively. We see the transition from the downstream to upstream regions at $x/\lsi \sim 400$, with the immediate upstream characterized with electromagnetic structures.  }
    \label{fig:shockStructure}
\end{figure*}

The late-time ($\omci t = 49.9$) shock structure is shown in Fig. \ref{fig:shockStructure}. For illustrative purposes, we show only the inner electromagnetic foreshock, containing the whistler waves, and the beginning of the outer electrostatic foreshock containing the EAWs. The full simulation box at this timestep extends to $\sim 2000 \lsi$. As explained in \cite{Bohdan2022} and \citetalias{Morris2022}, the EAWs here are not well captured because the simulation setup employs an out-of-plane field angle ($\phi=90^{\circ})$, with EAWs propagating in the direction of the upstream magnetic field. We therefore cannot see them so easily because they do not lie in the simulation plane. In contrast to \citetalias{Morris2022}, slightly ahead of the shock transition at $x/\lsi \gtrapprox 410$, we see large electromagnetic irregularities, which have developed from the oblique whistler instability. Their onset begins here at $\omci t \sim 10$, beyond the simulation time in \citetalias{Morris2022}, and they are associated with under-dense electron cavities (panel (a), also present in ions ) as well as magnetic- and electric-field turbulence (panels (b) and (c)). The physical scales of these cavities is of order $\lsi$, justifying the use of a larger transverse simulation box to allow a robust investigation of these phenomena. We further note that these magnetic and electric field inhomogeneities extend into the upstream beyond the shock ramp (for $x > 410 \lsi$) in the field profiles in panels (d) and (e), but decrease in strength \pjm{with} increasing distance from the shock. From these panels, where the field is averaged across the transverse direction of the simulation box, we see that $(B/B_0)^2$ approaches unity more quickly than $(E/E_0)^2$, indicating the end of the inner electromagnetic foreshock and the beginning of the outer electrostatic foreshock, the latter of which is the subject of \citetalias{Morris2022}. 

In \citet{Bohdan2022}, it was demonstrated via means of periodic boundary condition simulations that the differences in the inner and outer foreshocks can be explained by differences in the reflected electron populations that excite them. The \mpo{latter}, which leads to the excitation of electrostatic EAWs, has \mpo{in comparison to the inner foreshock} a reflected electron beam density a factor of 10 lower. Furthermore, the thermal spread of this electron beam is approximately \mpo{25\% lower than that of} the inner electromagnetic foreshock. These discrepancies are enough such that in the inner foreshock the electromagnetic oblique whistler instability is the dominant excited instability, as opposed to the electron acoustic instability which is prevalent in the outer regions.

The analysis presented in \citetalias{Morris2022} focused on the behavior of upstream electrons within the outer electrostatic foreshock. In the remainder of this paper, we focus on the inner electromagnetic foreshock, focusing on the properties of the waves, how they affect upstream electrons, and whether they can lead to electron pre-acceleration.

\subsection{Shock Reflection Rate}

\begin{figure}
    \centering
    \includegraphics[width=\columnwidth]{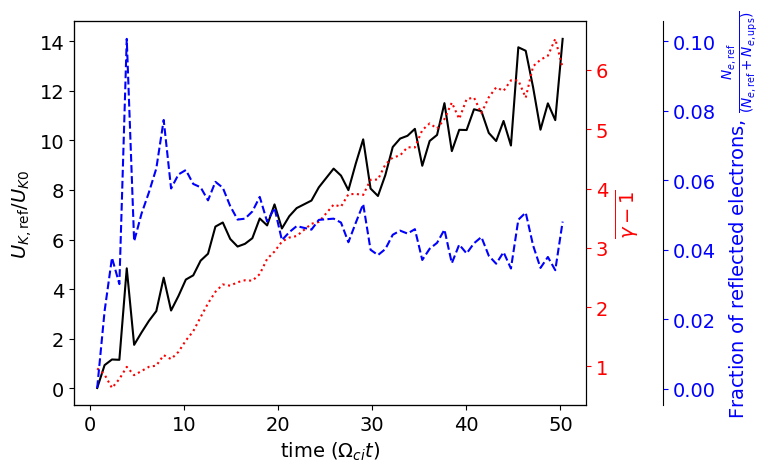}
    \caption{The reflection rate, $R = N_{e, \rm{ref}}/(N_{e, \rm{ref}}+ N_{e, \rm{ups}})$ where $N_{e, \rm{ref}}$ is the number of reflected electrons and $ N_{e, \rm{ups}}$ the number of upstream electrons that have not yet reached the shock, (blue dashed line), mean value of $\overline{\gamma -1}$ for the reflected electrons (red dotted line, simulation frame) and normalized kinetic energy contained within the reflected electron beam (black line). The latter quantity is normalized relative to the upstream kinetic energy such that $U_{K, \rm{ref}}/U_{K0} = R\overline{\gamma_{\rm ref}-1}/(\gamma_{\rm{ups}}-1)$ for reflection rate $R$ and $\gamma_{\rm ups} = (1-(v_{\rm ups}/c)^2)^{-1/2}$.  Quantities are calculated immediately ahead of the shock ramp between $\xsh + 2\lsi \leq x \leq \xsh + 10 \lsi$. Although the reflection rate falls with the onset of whistler waves at $\omci t \sim 10$, the overall energy density increases as the reflected electrons are more energetic on average. }
    \label{fig:refRate}
\end{figure}

We first quantify the energy content in the reflected electrons which are essential to excite the electromagnetic oblique whistler instability. We do so by first estimating the shock location, $\xsh$, taken as the location where $n_i/n_{0} = 2$ for ion number density $n_i$ and the subscript 0 denotes the far upstream value. We define this based on ion density as that of the ions reflected at the shock is around $10^{-4}n_{0}$ and is more stable in the foreshock relative to the electron number densities (due to higher electron reflection rates) and electromagnetic field amplitudes, which are disturbed by the whistler waves. In practice, this definition places $\xsh$ on the shock ramp, so we measure the reflection rate in the region defined by $\xsh + 2\lsi \leq x \leq \xsh + 10 \lsi$ to ensure we are measuring it for a region within the electron foreshock. Note that the parameters of the region of interest do not affect the presented results, so long as it resides within the electron foreshock. The chosen fixed region nevertheless lies close to the shock, and at late times is completely occupied by whistler waves, thus enables us to measure if they have any tangible effect on the reflection rate. 

\pjm{As measured in the simulation frame, the electron spectrum in the defined upstream region has a double peak structure in $\Gamma-1$, where the low energy peak corresponds to the thermal population moving with the upstream bulk flow and the high energy peak corresponds to reflected electrons. As in} \citetalias[][Fig. 4]{Morris2022}, \pjm{we use the local minimum between these peaks to distinguish between reflected and upstream electrons. At each timestep, reflected electrons are those within the defined region where $\Gamma-1$ exceeds the value for which ${\rm d} (N_e(\Gamma-1))/ {\rm d}(\Gamma-1) = 0$ and ${\rm d}^2 (N_e(\Gamma-1))/ {\rm d}(\Gamma-1)^2 > 0$ in the electron spectra. Those with $\Gamma-1$ lower than this threshold are considered to be upstream electrons traveling with the bulk flow of the incoming plasma. Note that this definition has no dependency on the direction in which the electron is traveling, thus an energetic reflected electron that is re-directed towards the shock is still considered reflected.} 

Fig. \ref{fig:refRate} indicates that \pjm{the onset of whistlers} may indeed \pjm{affect the reflection rate}. The blue dashed line shows that the reflection rate falls after the onset of whistler waves at around $\omci t \sim 10$, but is relatively stable at around $4\%$ for $\omci t > 30$. During the simulation, despite variations and a slow decline in the reflection rate, the energy density of the reflected beam increases at a roughly linear rate as indicated by the black solid line in Fig. \ref{fig:refRate}. \pjm{Possible causes of this include more efficient acceleration of reflected particles or an acceleration region which has a size that increases with time.} The red-dotted line shows that this can be explained by the fact that the mean Lorentz factor of reflected electrons also increases approximately linearly with time. 

We have verified that the reflection rate calculation is robust to our choice of region. The changes in reflection rate for the region shown in Fig. \ref{fig:refRate} occur further upstream in the same manner, although with a time-lag as it takes the reflected electrons longer to travel upstream along $\vec{B_0}$ and reach those regions. Accordingly, the energy density of the reflected electron beam increases throughout the simulation and reaches the threshold value to excite whistlers further from $\xsh$ as the simulation progresses, hence explaining why the size of the whistler region increases with simulation run-time. 


\section{Whistler Waves} \label{sec:Whistler}

\subsection{Wave Properties/Summary of Bohdan 2022}

A study of the instabilities driving the waves that arise in the electron foreshock was undertaken in \citet{Bohdan2022}, with results based on the same large-scale simulation that is presented here. In this earlier work, the electromagnetic waves present in the inner foreshock that we focus on in this paper were subject to a linear dispersion analysis, where it was established that they arise as a result of the oblique whistler instability. Evidence for this came from the fact that the waves in question have approximately the same parallel and perpendicular wavenumbers as well as growth rate as predicted by linear theory for the fastest-growing oblique whistler mode. Defining the wavevectors parallel and perpendicular to the upstream magnetic field as $\kpar$ and $\kperp$, respectively, and noting that the oblique whistler instability is excited by a beam of electrons moving parallel to the upstream flow with velocity $v_b$, we summarize the results of \citet{Bohdan2022} as: 

\begin{enumerate}

    \item The dependence of the perpendicular wave number, $\kperp$, of the fastest growing oblique whistler mode on the parallel velocity of the reflected electron beam is weak. For the parameters used here, the peak growth rate occurs at $\kperp \lse \approx 0.2$.
    \item The excited waves are in resonance with electrons reflected at the shock. Hence, $\kpar$ is extremely sensitive to $\vpar$, with the $l^{\rm th}$ order gyroresonance given by,
    \begin{equation}
        \omres = k_\parallel v_{b} - \ell~\frac{|\omce|}{\gamma_b}, \hspace{1cm} l \geq 1
        \label{eq:resonance}
    \end{equation}
    for beam Lorentz factor $\gamma_b$ and electron gyrofrequency $\omce$. 
    \item The beam can only excite fluctuations with sufficiently small phase speed ($\mathcal{O}(10^{-3}c$)), i.e. waves with angular frequencies satisfying  $\varpi_W < \omres$. 
    \item Decreasing the beam number density of the reflected electrons results in a smaller growth rate of the whistler mode.
    
\end{enumerate}

From this latter point, and from \citet[][Fig. 9]{Bohdan2022}, we note that the dependence of $\omega$ on $k$ for the whistlers is approximately linear, and hence the group velocity, $v_g$, is of the same order of magnitude as the phase velocity, meaning $v_g << v_{\rm up}$. Accordingly, we see the linear structures co-move with the upstream bulk flow when viewed in the frame of reference of our simulation.

\begin{figure}
    \centering
    \includegraphics[width=\columnwidth]{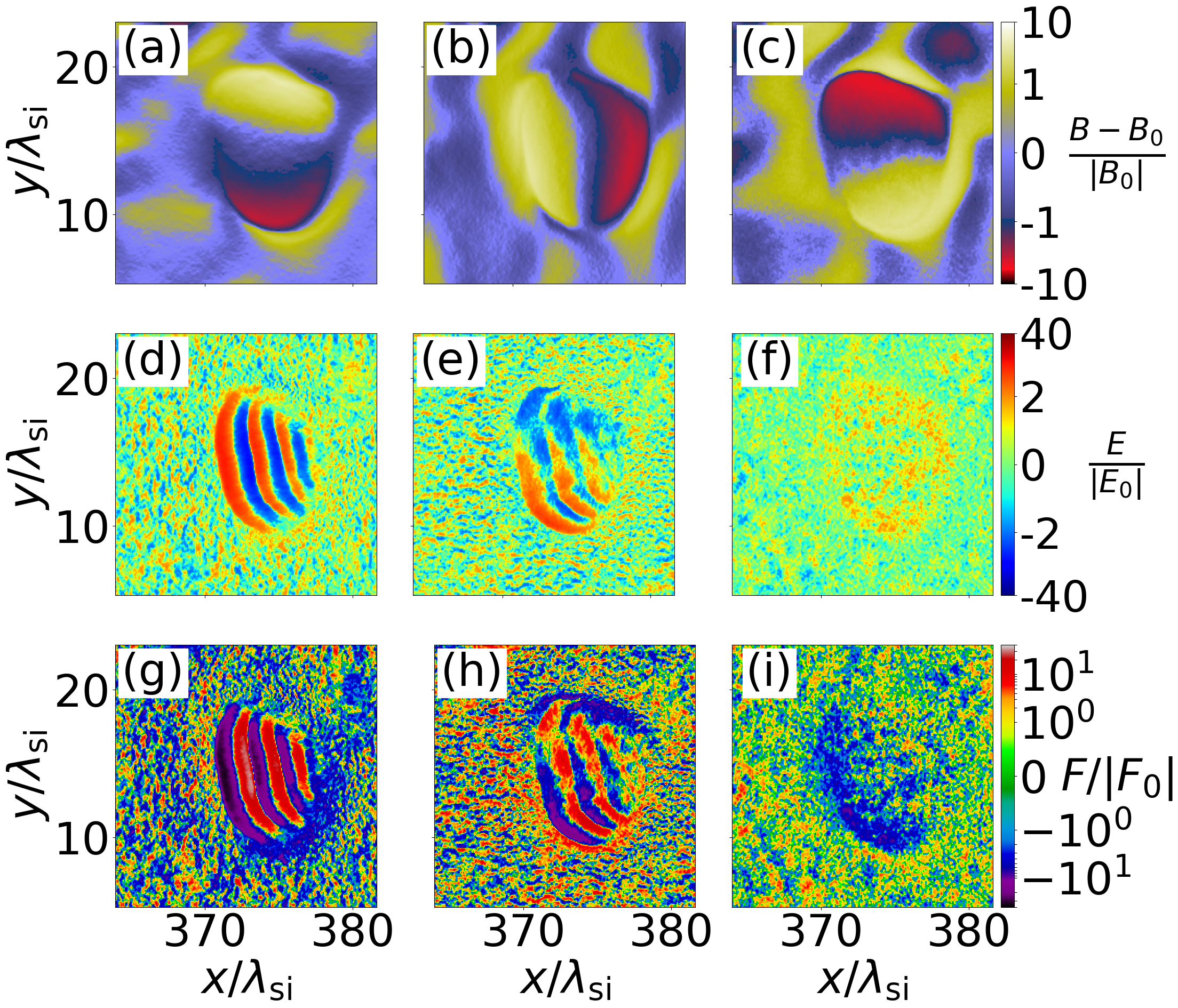}
    \caption{The structure of a prominent whistler wave packet present in the electromagnetic foreshock at $\omci t = 42.1$ is shown as measured in the upstream rest frame. For visual clarity, quantities are normalized relative to background simulation-frame field values (note there is no motional $E$-field in this frame). Panels (a), (b) and (c) show $B_x$, $B_y$ and $B_z$, panels (d), (e) and (f) show $E_x$, $E_y$ and $E_z$.  Panels (g), (h) and (i) show the Lorentz force for a test particle moving at $\vec{v} = (-v_{e,\rm{th}}/\sqrt{3})\cdot(1,1,1)$, for electron thermal velocity $v_{e,\rm{th}}$. Here, the Lorentz force components are normalized to the modulus of $F_0 = e v_{e,\rm{th}}B_0$. }
    \label{fig:Wfields}
\end{figure}

\subsection{Non-linear Structures}

Although the behavior of the initial whistler wave structure can be appropriately described by linear analysis, at later times in the simulations they develop into non-linear wave packets, with a complex internal structure. In this section we outline the structural evolution of the whistler waves into non-linear wave packets as they propagate towards the shock from the upstream.

Fig. \ref{fig:Wfields} shows the electromagnetic field structure of a particularly prominent non-linear structure (developing from a whistler wave) occurring in the inner electromagnetic foreshock of the simulation at $\omci t = 42.9$. Quantities have been measured in the upstream rest frame, which removes the large scale motional electric field, thus the electric field structure displayed is dominated by that associated with the non-linear structures. The top row illustrates the magnetic field structure, with the middle row showing the electric field. From left to right the figure shows the x-, y-, and z- components. We note that while all three magnetic field components have roughly similar peak magnitudes, the maximum absolute values of the $E_x$ and $E_y$ components are around an order of magnitude higher than for $E_z$. 

\begin{figure}
    \centering
    \includegraphics[width=\columnwidth]{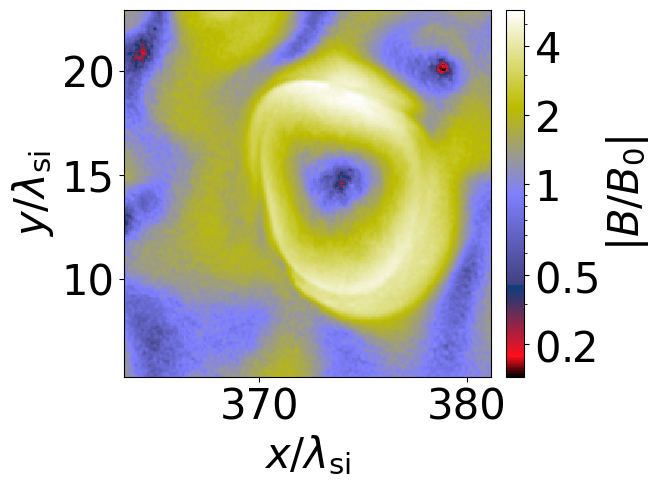}
    \caption{The structure of $|B|/|B_0|$ for the non-linear structures. The magnetic energy density is strongest in a ring-like shape encircling the structure, with the field strength declining towards the center and radially outside of the area of maximum strength.}
    \label{fig:cavityB}
\end{figure}

We further note that the characteristic size of the whistlers increases as they approach the shock, which is consistent with our previous studies in \cite{Bohdan2022}. \pjm{It is from within these waves at late times that highly non-linear structures develop.} From panels (a) - (c) \mpo{of Fig.~\ref{fig:Wfields}}, we see that the $B$-fields are in general strongest in magnitude at the edge of the whistlers, and progressively weaker towards the central region, such that in 2D space the value of $|B/B_0|$ reaches a maximum in a ring shape encircling the \mpo{nonlinear wave structure}, as depicted in Fig. \ref{fig:cavityB}. The range of spatial radii from $1-5 \lsi$ of these structures is similar to that of short large amplitude magnetic structures (SLAMS) \citep{Mann1995, 2020ApJ...898..121W}, which have been detected in the bow shocks of the Earth \citep{1994JGR....9913315M}, Venus \citep{2020ApJ...898..121W}, and Jupiter \citep{1993P&SS...41..851T}. \pjm{However,} in contrast to SLAMs, where the density is amplified by a factor of a few relative to the upstream plasma, we see from Fig. \ref{fig:shockStructure} that the non-linear structures discussed here are associated with under-dense cavities, with under-densities as low as $n/n_0 \approx 0.15$. \pjm{We further note that observational evidence for whistler-mode induced structures has been provided by analysing satellite data from the Magnetospheric Multiscale mission \citep{2021arXiv211114832H,2022arXiv221105398S}. These observational data apply to the Earth's bow shock, thus the parameters used here are not consistent with or supernova remnant based simulation.} 


To comprehend any pre-acceleration that arises consequentially from the interactions of electrons with whistler waves, we must understand the forces they experience when interacting with them. In general, an electron with charge $q_e$ immersed in both electric (${\boldsymbol E}$) and magnetic (${\boldsymbol B}$) fields experiences a Lorentz force, ${\boldsymbol F}$,  which is defined as,

\begin{equation}
    {\boldsymbol F} = q_e({\boldsymbol E} + {\boldsymbol v} \times {\boldsymbol B}) 
    \label{eq:florentz}
\end{equation}
\pjmii{where bold quantities represent vectors with $x$-, $y$- and $z$- components in Cartesian space. From} Eqn. \ref{eq:florentz}, the velocity of an electron, both in terms of direction and magnitude, can influence the resulting behavior when interacting with a region of strong electromagnetic fields, such as those within the non-linear structures.

We trace a sample of 10,000 upstream electrons to probe any interactions with the whistler waves and non-linear structures. These electrons are selected randomly from the upstream population at $\omci t = 30$ from a region between $360 \lsi$ and $410 \lsi$ ahead of the shock, and traced for the remainder of the simulation (over 20 $\omci^{-1}$). Such a sample is representative of the global upstream population, and our sampled region permits an adequate duration for them to pass through the foreshock and interact with the shock itself.

\subsection{Interaction of upstream (bulk flow) electrons with whistlers}

We have already established that the non-linear structures have phase and group velocities of $\mathcal{O}(10^{-3}c)$. Because of this, they are quasi-stationary when viewed in the upstream rest frame, and any force components arising from their motion relative to the upstream bulk plasma is negligible. For these reasons, the upstream rest frame is appropriate to analysis interactions between the structures and upstream electrons.  

In this frame, the upstream electron population is thermal, with a most probable speed of $\veth = 0.044c$. This corresponds to an electron gyroradius of $r_{ge} = \veth \wce^{-1} = 5.7 \Delta \approx \lsi/10$, which is around two orders of magnitude smaller than the size of the largest non-linear structures such as that shown in Fig. \ref{fig:Wfields}. As $B_y = 0$ and $B_z > B_x$, the $B_y$ terms in \pjmii{the Lorentz force (see }Eqn. \ref{eq:florentz}\pjmii{)} can be neglected. The colormap of panels (g), (h) and (i) in Fig. \ref{fig:Wfields} show the Cartesian component of the Lorentz force as measured in the upstream rest frame that would be experienced by a test electron moving with the upstream bulk flow such that its velocity is given by $\vec{v}_{\rm{test}} = -\veth/\sqrt{3}(1, 1, 1)$. We choose Cartesian velocities of $v_{\rm{test}}$ such that the thermal velocity magnitude is divided equally between them.

Thermal upstream electrons that are spatially coincident with the growing non-linear structure will experience forces according to Eqn. \ref{eq:florentz}. In the $x$-direction, the direction of $E_x$ oscillates and Fig. \ref{fig:Wfields} shows that this oscillation dominates the structure of $F_x$. This \pjm{in} combination with gyroradii typically much smaller than the radial extend of the structure ensures that the electrons will in general remain co-moving with the non-linear structure in the $x$-direction. Additionally, the force in the $z$-direction is comparatively weak relative to other components.

The interaction of upstream co-spatial electrons with the non-linear structures is much more interesting in the $y$-component. From Eqn. \ref{eq:florentz}, $F_y = q_e (E_y + v_zB_x - v_xB_z)$. Firstly, when considering cool thermal upstream electrons, the $q_e E_y$ term dominates, thus the overall Lorentz force directs them outwards and away from the non-linear structure, helping to carve out low density cavities. To understand the behavior of hotter upstream electrons we need to account for the signs of the three terms that constitute $F_y$ and the field geometry shown in Fig. \ref{fig:Wfields}. Here, we note that the cross terms are in the same direction as the $q_e E_y$ term if they are both negative, as is the case for our test particle. In fact, this scenario is both plausible and likely as from Fig. \ref{fig:Wfields} $B_x$ and $B_z$ are generally diametrically opposed (thus $v_x$ and $v_z$ share the same sign, and are positive and negative each for half of one gyration period). In the negative case, as for cool electrons the Lorentz force is again directed outwards. However, when $v_x$ and $v_z$ are both positive, the magnitude of this outwards force reduces. Despite this, the overwhelming majority of upstream electrons are too cool for the force to ever point inwards, with the overall force away from the structure center when averaged over the electron gyro-period.  

\begin{figure}
    \centering
    \includegraphics[width=\columnwidth]{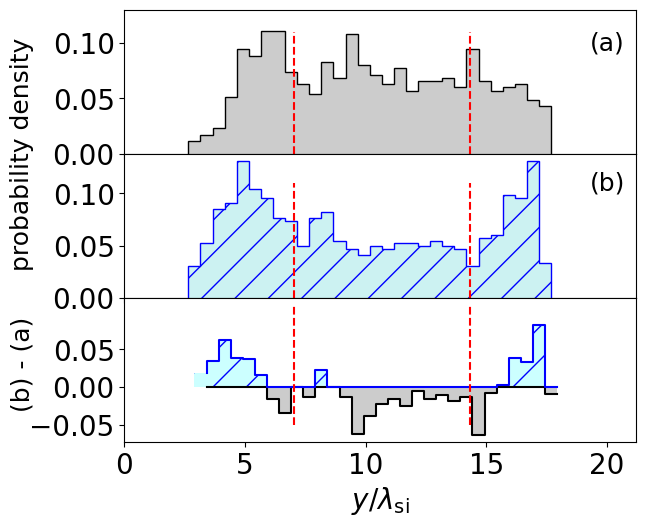}
    \caption{ Probability distribution along the transverse direction before and after the wave shown in Figs \ref{fig:Wfields} and \ref{fig:cavityB}. Due to the shape and strength of the $E_y$ component of the whistlers, electrons spatially coincident with the top half experience an upwards force, while those approaching the lower half an downwards force. In these plots, we see the initial distribution (top, at $\omci t = 41.3$) and distribution after (lower panel, at $\omci t = 42.4$), which is now bimodal. The vertical dashed red lines indicate the radial extent of the non-linear structures.}
    \label{fig:histComp}
\end{figure}

In general, as shown in panel (h) of Fig. \ref{fig:Wfields}, there is a net force away from the center of the non-linear structures on co-spatial upstream electrons.  Using a sub-set of our traced electrons that are located within $2 \lsi$ of the radial extent of the structure, we see that the direction of the Lorentz force leads to a bi-modal distribution of these electrons, as shown in Fig. \ref{fig:histComp}. Here, we see a slight preference for the electrons to be present beyond the radial extent of the structure, as opposed to centrally within it. This effect is only noticeable for non-linear waves with particularly large amplitudes, and in general it preferentially expels relatively cooler electrons.

Crucially, we note that for upstream electrons the force associated with the non-linear structures is small, and not generally towards the wave center. This means that upstream electrons are not likely to resonate with these waves, making the scenario where upstream electrons are trapped in the non-linear structures that have developed within the whistler potential highly unlikely. 

\subsection{Interaction of Reflected Electrons with Non-linear Structures}

\begin{figure*}
    \centering
    \includegraphics[width=\textwidth]{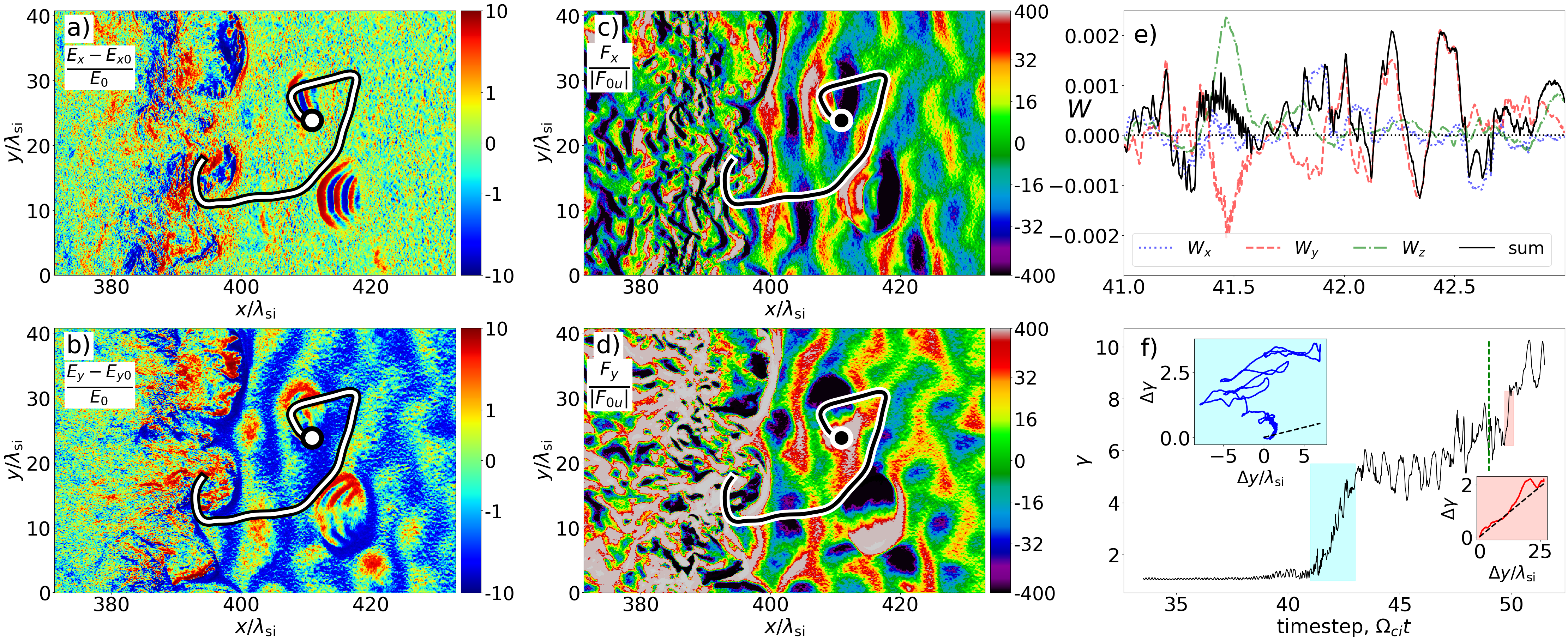}
    \caption{Panels (a) - (d) show the trajectory of a reflected electron though the simulation, with the current location at $\omci t = 49$ marked by the circle center. Its path over the previous $\omci^{-1}$ is shown by the black and white line. The background in panels (a) - (b) show fluctuations in $E_x$ and $E_y$, while (c) and (d) show the Lorentz force components $F_x$, and $F_y$, \pjm{using the particle velocity at $\omci t = 49$ of $v/c = (0.47, 0.04, 0.86)$} normalized relative to $F_1 = e v_{\rm up}B_0$, respectively. We see that these force components are now directed towards the wave center. Panel (e) shows the work done on the electron by all Cartesian components of the electric field in its own rest frame. The interval shown here corresponds to the cyan-shaded area of panel (f) which shows the electron Lorentz factor as a function of time. Here, the dashed green line shows the current time of $\omci t = 49$. The solid lines within the cyan and salmon colored panels show the change in Lorentz factor vs $\Delta y$ for the color-indicated regions, and the analytically expected $\Delta \gamma$ vs $\Delta y$ from SDA is plotted with a black dashed line. For aesthetic purposes, the color bar label has been moved inside the figure for panels (a)-(d). We see two prominent kinks in the particle trajectory. The first of these occurs at $\omci t = 48.44$, which is caused by the electron deflecting around the lower half of a non-linear structure (note that the structure centered at (415, 12) here is not responsible, as it was further upstream at this timestep). The second kink occurs at $\omci t = 48.75$, at which point the electron is trapped by the non-linear structure we see it co-move with afterwards.  }
    \label{fig:trPar}
\end{figure*}

In general, as measured in the upstream reference frame, reflected electrons have positive values for at least two Cartesian velocity components. A positive $v_z$ is needed as they travel along the upstream magnetic field lines, of which the strongest component lies along $\hat{z}$. An additional magnetic field component in $\hat{x}$, in combination with the fact that a positive $v_x$ is required so the electron can outrun the shock (such that $v_x > v_{\rm{sh}}\cos \thbn = v_{\rm{sh}}/2$) ensures $v_x$ is also positive. If this latter criterion is not met, it cannot be reflected. 

When considering the $v_y$ component, we first note that reflected electrons require some pre-acceleration to be reflected from the shock. These mechanism tend to provide acceleration due to work done by the motional electric field, which here lies in the $-\hat{y}$ direction, leading to electron acceleration in the $\hat{y}$ direction by virtue of their negative charges. This velocity component is purely gyrational, so oscillates around zero at the electron gyrofrequency. 

From Eqn. \ref{eq:florentz}, we see that this changes the forces a reflected electron experiences during an encounter with a whistler wave relative to an upstream electron. Fig. \ref{fig:trPar} shows such an interaction in the simulation frame. Panels (a) and (b) show fluctuations in $E_x$ and $E_y$ (to visualize the other field components of the whistler see Fig. \ref{fig:Wfields}), while (c) and (d) show $F_x$ and $F_y$, respectively. Here, they are normalized relative to modulus of $F_{1} = e v_{\rm{up}}B_0$. Panel (e) shows the Cartesian components of acceleration felt by the electron in its own rest frame. (f) shows the Lorentz factor of the electron. The red dashed line in this panel indicates the timestep corresponding to the images. For panels (a) - (d) the trajectory of the electron during the previous ion gyroperiod is shown by the black-and-white lines.

Immediately we see that the trajectory of the reflected electron is deflected away from the direction of $\vec{B_0}$ and is influenced by the presence of the developed non-linear structures. Furthermore, the electron appears to have been trapped by the (prominent) upper structure located at $(x/\lsi, y/\lsi) = (410, 27)$. Crucially we can determine from Eqn. \ref{eq:florentz} that, in contrast to upstream electrons, the $F_x$ and $F_y$ forces are now directed towards the center of the non-linear structure. If we consider the $y$-direction in the simulation frame, in comparison to an interaction with an upstream electron (where $v_x < 0$), the $v_x B_z$ term is now aligned with the $E_y$ term, with each of these pointing inwards. If all three velocity components are positive, no matter how the electron approaches the whistler, all three terms in $F_y$ (Eqn. \ref{eq:florentz}) point inwards, enabling trapping. The three Cartesian components of the Lorentz force generally point inwards for reflected electrons. As measured in the upstream rest frame and relative to the upstream magnetic field, \pjm{typical Larmor radii for reflected electrons are around 1-5 $\lsi$, but can be up to $\approx 10 \lsi$}, which from Fig. \ref{fig:cavityB} are compressed by a factor of around 4 when encountering the strong magnetic field associated with a non-linear structure, thus reflected electrons can typically be contained within them.

\subsection{Interaction Probability}

We can estimate the probability that a reflected electron will interact with a \pjm{non-linear structure} by considering the path taken by such a particle. The whistler waves and resulting non-linear structures move with group velocity $v_g << v_{\rm{up}}$, hence we perform this calculation in the upstream rest frame using the \pjm{approximation} that the non-linear structures are stationary. We use primed quantities here to represent the upstream rest frame, and further assume the size of the whistler-containing electromagnetic foreshock \pjm{from which the non-linear structures derive} to be a constant size of $x_w'$. Primed quantities with the subscript $\parallel$ and $\bot$ refer to components measured parallel or perpendicular to the magnetic field vector in the upstream rest frame, respectively. 

Since the reflected electrons are gyrating, we can consider the path length to be a sum of 2 components. These consist of a linear component, $R_{\rm{lin}}'$ as a result of the path of reflected electrons following the large-scale upstream magnetic field structure, and an oscillatory component, $R_{\rm{osc}}'$, as a result of the gyration around the magnetic field. The total path length can be considered to be, 
\begin{equation}
    \rpathp = \sqrt{ \rlinp^2 + \roscp^2}
    \label{eq:rp1}
\end{equation}
where $\roscp$ dominates if the electron Larmor radius is significantly larger than $x_w'$. Otheriwse, $\rpathp \approx x_w'$.

$\rlinp$ is simply the size of the whistler-containing region, such that $\rlinp = x_w'$. To compute the time an electron takes to traverse it, we consider that they travel in the direction of the upstream rest frame magnetic field, $\vec{B_0}'$ which now makes an angle $\thbn'$ with the simulation plane. We therefore only see their propagation projected onto the simulation ($xy$) plane, with the time taken to traverse $\rlinp=x_w'$ defined as,
\begin{equation}
    t_{\rm{lin}}' = \frac{x_w'}{\vpar' \cos \thbn'},
\end{equation}
where the $\cos \thbn'$ term provides the necessary path correction to compensate for the inclination of the upstream magnetic field with respect to the simulation plane.

During time $\tlinp$, assuming it does not interact with any non-linear structures, a reflected electron completes $\tlinp/\tau_{ge}$ oscillations, where $\tau_{ge} = 2\pi r_{ge}/\vperp$ is the electron gyro-period. As a result of our magnetic field orientation with respect to simulation plane, the shape made by the gyrational orbit of the electron on the simulation plane is elliptical. On account of $B_y' =0$, the size of the semi-major axis is $a=r_{ge}$, where $r_{ge}$ is the electron gyroradius. The semi-minor axis appears contracted in the direction of motion, such that its value is given by $b = r_{ge} \sin \thbn'$. For the sake of presenting a more easily interpretable solution, we approximate the perimeter of the elliptical path, $p$, as,
\begin{equation}
    p \approx 2 \pi \sqrt{ \frac{a^2+b^2}{2} } = \sqrt{2} \pi r_{ge} \sqrt{1 + \sin^2 \thbn'},
\end{equation}
which is typically accurate to better than 5\% assuming the ellipse is not too elongated \citep{1902Natur..66..174M}. Other, more accurate, approximations can be found in the literature \citep{ramanujan2015collected}. The total gyrational path length is therefore given by $\roscp = p \tlinp /\tau_{ge}$.

Combining these, we can rewrite Eqn. \ref{eq:rp1} as, 
\begin{equation}
    \rpathp = x_w' \left( 1+ \frac{\vperp'}{\sqrt{2}\vpar'} \sqrt{ \frac{1+\sin^2 \thbn'}{1-\sin^2 \thbn'} } \right) ,
    \label{eq:rpath}
\end{equation}
where this equation must also satisfy the reflection constraint that $\vpar' \cos \thbn' \geq v_{\rm{sh}}$ to remain valid.

The probability of interaction with a non-linear structure follows,
\begin{align}
  \begin{split}
    P_{\rm interact} &\approx  n_{\rm{nls}} \sigma_{\rm{nls}} R_{\rm path}'\\
    &= n_{\rm{nls}} \pi r_{\rm{nls}}'^2 x_w' \left( 1 + \frac{1}{2} \left(\frac{\vperp'}{\vpar'}\right)^2  \left( \frac{1+\sin^2 \thbn'}{1-\sin^2 \thbn'} \right)   \right)^{1/2},
    \label{eq:Pint}
  \end{split}
\end{align}
for \pjm{non-linear structure} number density $n_{\rm{nls}}$ and cross-sectional area $\sigma_{\rm{nls}} = \pi \pjm{r_{\rm{nls}} }'^2$ as measured in the upstream rest frame. We can interpret our results within the context of Eqn. \ref{eq:Pint}.

Firstly, we recover the intuitively expected results that both a larger number density of non-linear structures and their cross-sectional area linearly increase the interaction probability. In the limit $\vpar'  \rightarrow c$, $\vperp' \rightarrow 0$ to prevent the electron becoming superluminal, and we recover the expected solution that $R_{\rm path}' = x_w'$ (the same is true if $\vperp'$ is small). Eqn. \ref{eq:Pint} also recovers the expected solution that the path length approaches $\infty$ in the case of a perpendicular shock where $\thbn' \rightarrow 90^{\circ}$ because the reflected particles are unable to escape upstream.

More significantly, we note that the second term in Eqn. \ref{eq:Pint} indicates that the probability of interaction is proportional to $\vperp$, but inversely proportional to $\vpar$. One may conclude that this may favor reflected electrons with $\vperp' >> \vpar'$, however this is not accurate. For a more realistic picture, we must again consider that for a reflected electron to outrun the shock it must satisfy $\vpar' \cos \thbn' \geq v_{\rm{sh}}$. This, in addition to the constraint that $v'^2 = \vperp'^2 + \vpar'^2 \leq c$, restricts the value of $\vperp'$ to be $\vperp' \leq  0.848 c$.

The electron shown in Fig. \ref{fig:trPar} interacts with a non-linear structure, and becomes trapped. During this time, it is carried towards the shock as it is unable to escape, until the wave `breaks' when encountering the shock ramp. We note that the Larmor radius of reflected electrons that are trapped by whistler waves need to be smaller in size relative to the whistler wave. \pjm{Typically, when measured relative to the upstream magnetic field, reflected electrons have gyroradii in the range of $1-5 \lsi$, though can approach $10 \lsi$. This comparison means the ratio of reflected electron gyroradii to the radius of a non-linear structure lies in the range $0.1 - 1$, as the non-linear structures measure $\approx 5 \lsi$ radially. Although the magnetic field amplification associated with these structures will contract electron gyroradii that encounter them, this may not be sufficient to allow the non-linear structures to trap the most energetic reflected electrons.} In reality, we might expect the probability of trapping to fall off exponentially as the gyroradius approaches the radius \pjm{of the non-linear structure, $r_{\rm{nls}}$,} such that $P_{\rm trap} \propto \exp(-r_{ge}/r_{\rm{nls}})$. 

With a simplifying assumption that reflected electrons would move with a constant $\vpar$ in a calm upstream region, we can calculate the most probable value by assuming $P_{\rm trap} \propto P_{\rm interact} \cdot \exp(-r_{ge}/r_{\rm{nls}})$. Numerically solving this equation and accounting for the aforementioned constraints, under our simulation setup we obtain a peak trapping probability for $\vperp' \approx 0.40 c$ and $\vpar' \approx 0.54c$, which corresponds to reflected electrons that are around 18 times as energetic as thermal electrons in the far upstream. 

\pjm{While this analysis has allowed us to estimate trapping conditions for reflected electrons, it only provides a snapshot over a small time-frame. 
More realistically, \citet{Bohdan2022} show that the size of the whistler region from which the non-linear structures arise from grows with time. By integrating the energy density of the reflected electrons up to the point that the whistler waves become detectable and extrapolating this to late times, \citet{Bohdan2022} estimate that the size of this region would reach a steady state at $\omci t \approx 125$. At this time, the size of the whistler containing region would extent to around $2000 \lsi$ ahead of the shock. Within the context of our analysis, from Eqn. \ref{eq:Pint} we would expect the trapping probability to increase up until $\omci t \approx 125$ for all reflected electrons with gyroradii small enough to be contained by the non-linear structures. Further upstream beyond this region, the energy density of the electron acoustic waves (which are not well captured in this out-of-plane simulation) that are the subject of  \citetalias{Morris2022} would dominate, hence the non-linear structures discussed here would cease to be important beyond this limit.}

\subsection{Stochastic Shock Drift Acceleration}

\begin{figure}
    \centering
    \includegraphics[width=\columnwidth]{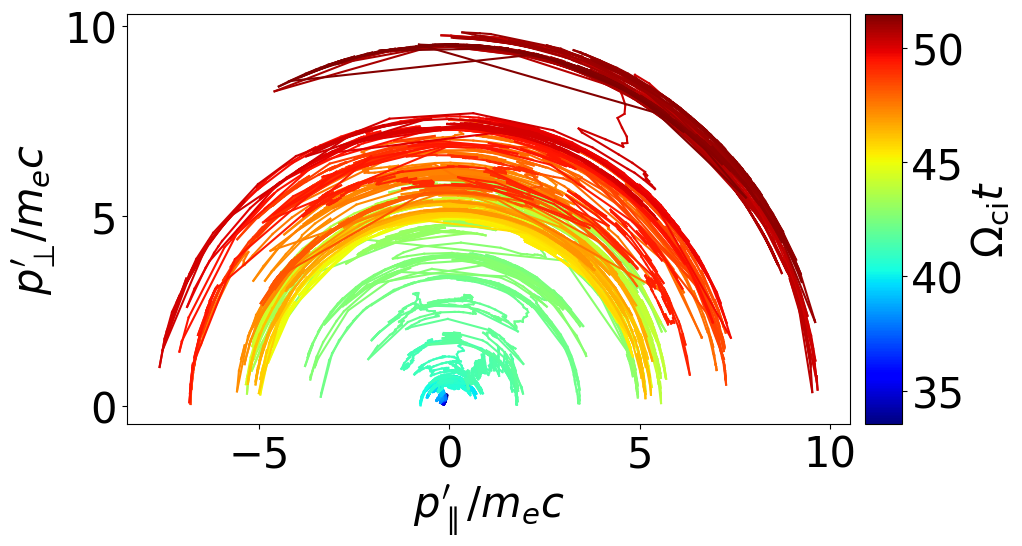}
    \caption{Plot showing perpendicular vs parallel momentum for the electron in Fig. \ref{fig:trPar}, as measured in the upstream rest frame relative to the local magnetic field. Arcs of constant momentum indicate pitch angle scattering, while increases in $\pperp'$ are suggestive of SDA. }
    \label{fig:ppplot}
\end{figure}

From Eqn. \ref{eq:Pint}, electrons with some perpendicular acceleration are more likely to be trapped by the non-linear structures, assuming that they have enough parallel acceleration to escape the shock front and under the condition that their gyroradii are smaller than the characteristic radius of the non-linear structures. Trapped electrons will be returned to shock where they may undergo further pre-acceleration. Identifying a mechanism that provided perpendicular acceleration can explain much about their behavior. 

Known mechanisms in this region that increase $\vperp$ include shock surfing acceleration (SSA) and stochastic shock drift acceleration (SSDA). The former of these processes is dependent on the presence of electrostatic Buneman waves, which are excited by a velocity difference between incoming electrons and reflected ions, which results in their production near the shock ramp \citep{Buneman1958,Gary1987}. In perpendicular shocks, ion gyration at the shock is sufficient to excite them in the shock foot \citep{Bohdan2019a}. In oblique shocks, the Buneman instability is strongly modified due to the presence of whistler waves and so the overall efficiency of SSA might be different compared to perpendicular shocks. Note that ions propagating back upstream cannot drive Buneman waves in the foreshock region since the ion reflection rate is too small, $n_{\rm ion, ref}/n_0 = 10^{-4}$, and the growth rate predicted for Buneman waves is over two orders of magnitude smaller than that for the whistler waves \citep{Bohdan2022}.

Another candidate acceleration mechanism is shock drift acceleration (SDA). The original theory of SDA indicated that it could efficiently accelerate charged particles \citep{1984JGR....89.8857W,1984AnGeo...2..449L}. It occurs if the electron gyrates close enough to the shock ramp such that part of its orbit overlaps the region with enhanced magnetic field, tightening its gyro-radius during these regions. The gradient in the magnetic field results in a drift analogous to $\nabla B$ drift with work done by, and in the direction of, the motional electric field, which is perpendicular to $\vec{B_0}$ by definition \citep{Ball2001}. 

Despite the efficient energization, \citet{Vandas2001} demonstrated that SDA alone is not efficient enough to account for the observed power-law spectrum and fluxes of accelerated electrons in astrophysical sources. Physically, this occurs because in the original SDA theory, candidate electrons are not confined to the shock transition region where acceleration occurs, thus limiting the efficiency of the mechanism. One such way of overcoming this impediment is to add pitch angle scattering, with electrons scattering off whistlers being observed in the Earth's bow shock \citep{Oka2017}. \citet{Katou2019} proposed the stochastic shock drift acceleration (SSDA) mechanism which incorporates pitch angle scattering into the SDA model, increasing the time in the acceleration region and accordingly the energy gain. Additional evidence of SSDA has been found in both in 3D PIC simulations \citep{Matsumoto2017} and in observations which support electron scattering by whistler waves \citep{Oka_2019}. This latter work concludes that the energization directly via the whistlers is low, but the electrons, as here, are confined within the acceleration region and become more energetic. This picture is consistent with our results.  Fig. \ref{fig:ppplot} plots $\pperp'$ vs $\ppar'$ (where the primes again represent the upstream rest frame) for the electron shown in Fig. \ref{fig:trPar}. We see increases in $\pperp'$, supporting SDA as the mechanism that provided the acceleration, and variations in pitch angle for constant $p'$, which are indicative of scattering \citep{Matsumoto2017,2021ApJ...915...18H}. The vertical red line at around $(\ppar', \pperp') = (5, 7.5)$ corresponds to the salmon pink region of Fig. \ref{fig:trPar} panel (f). This occurs when $\omci t \approx 49$, when the electron has been returned to the shock by the non-linear structure that had trapped it. Another signature of SSDA is that the change in energy is directly proportional to the motional electric field. For our magnetic field configuration electrons will drift in the $+\hat{y}$ direction \citep{1989JGR....9415367K}. We see from panel (f) that for the salmon-pink region, the predicted $\Delta \gamma$ (dashed black line) agrees closely with measured values (solid red line), verifying that SSDA is observed in our simulation.

\begin{figure}
    \centering
    \includegraphics[width=\columnwidth]{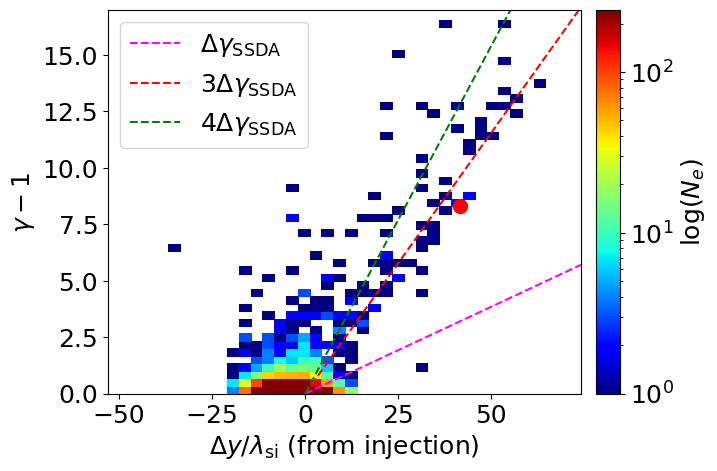}
    \caption{The change in Lorentz factor as a function of $\Delta y/\lsi$ at $\omci t = 51.5$ for the traced electrons. The negative average $\Delta y$ typically occurs for electrons that have passed into the downstream. The location of the electron shown in Fig. \ref{fig:trPar} is indicated by the red circle. Under the assumption that the work done to cause the acceleration comes from the motional electric field, SSDA predicts $\Delta \gamma_{\rm SSDA} \approx q_e E_{0y} \Delta y/(m_e c^2)$, thus the energy gain for the most energetic electrons is 3-4 times greater than can be accounted for by SSDA from the motional electric field alone. }
    \label{fig:SDAtrace}
\end{figure}

However, the presence of the non-linear structures further complicates this picture, and the acceleration of the most energetic electrons cannot be fully described by SSDA alone. Panel (f) in Fig. \ref{fig:trPar} indicates that this particular electron undergoes two periods of rapid and efficient acceleration, with relative quiescence in between. At around $\omci t = 41$, the electron first encounters the shock and is accelerated, with this region indicated by the cyan panel and corresponding to the color-matched sub-panel. The phase of constant energy corresponds to the electron traveling upstream, and includes the time-period when it is trapped by the non-linear structure, indicating the primary role of such structures is to keep electrons confined to the acceleration region.  The cyan panel shows the measured $\Delta \gamma$ (blue solid line) and analytical $\Delta \gamma$ (black dashed line) as a function of $\Delta y$ for the first acceleration period. Although there are incidences where the energization rate can be attributed to SSDA, $\Delta \gamma$ is shown to also increase for $\Delta y \leq 0$. We see that this behavior is also evident in the subset of the most energetic traced electrons. Fig. \ref{fig:SDAtrace} indicates that they generally have a linear relationship between $\Delta \gamma$ and $\Delta y$, as for the trace electron shown in Fig. \ref{fig:trPar} is shown on Fig. \ref{fig:SDAtrace} by the red circle. However, on average, the acceleration is between three and four times more efficient than can be accounted from SSDA via the motional electric field alone, with this prediction indicated by the purple dashed line. This requires further investigation.

To explain this, we compute the total work done, $W$, by electric field on the electron in its own rest frame. This is a sum of the Cartesian components, such that $W = W_x + W_y + W_z$, with these displayed in Fig. \ref{fig:trPar} panel (e) for the interval $\omce t = 41 - 43$ which corresponds to the cyan area in panel (f). Noting that $(\Delta y/\lsi, \Delta \gamma) = (0,0)$ on the latter plot corresponds to $\omci t = 41$ on panel (e), we see that initially when $W_y \leq 0$ the change in $y$-coordinate is $\Delta y \leq 0$. Closer to $\omci t = 43$, $W_y$ becomes positive, in which regions we see acceleration consistent with SSDA (the blue solid line is quasi parallel to the black dashed line in panel (f), cyan background). However, the important distinction to SSDA is that the overall acceleration during this period is typically greater than zero. We assess the relative importance of each component by computing the mean work done across this time interval. For this electron, the $W_z$ component is particularly strong, and has a relative contribution to the overall particle energy change that is comparable to that from $W_y$, with $W_z \approx 2.1 W_y$. $W_x$ is weaker at $W_x \approx 0.17 W_y$. Typically for reflected electrons, we see similar levels of work from $W_z$ and $W_y$ with a smaller $W_x$, unlike in SSDA where we would expect $W_y$ to drive the acceleration exclusively. The reason we see more efficient acceleration overall relative to SSDA is therefore because all Cartesian components can in principle contribute, thus we would expect $\Delta \gamma / \Delta \gamma_{\rm SSDA} \geq 2$ on account of the comparable $W_y$ and $W_z$, which is consistent with what we show in Fig. \ref{fig:SDAtrace}.

The precise details of the underlying microphysics of the acceleration are likely a consequence of the complex non-linear structures arriving at the shock. Indeed, previous studies have shown that changes in local conditions can lead to more efficient electron acceleration \citep{2021ApJ...919...97K}. For the electron considered in Fig. \ref{fig:trPar}, we compute the dot-product of $\vec{B}_0$ and the time averaged local magnetic field from $\omci t = 41.4$ to $\omci t = 41.5$, when $W_z$ dominates the work done by the electric field in the particle rest frame (panel (f)). The cosine of the angle between then is about $ 0.9$, whereas for the pure SDA consistent region (i.e. pink shaded region of Fig. \ref{fig:trPar}(f) ) this number is $\approx 1$. This means that the local magnetic field in these regions on average subtends an angle with the upstream field of around $26^{\circ}$ during this period. As the work done during acceleration via SDA occurs via purely perpendicular electric fields, this changing of the local field orientation permits acceleration in additional directions. Indeed, the non-linear structures arriving at the shock also perturbate the local electric fields, with $E_y/E_{0y}$ ranging from $\pm 15$ in the cyan region of Fig. \ref{fig:trPar}f, yet averaging at $E_y/E_{0y} \approx 1$. The overall acceleration is therefore highly sensitive on local values and orientations of the electromagnetic fields.

\subsection{Pitch Angle Distribution}

On the one hand, an electron with a larger $\vperp'$ will have a longer path length via Eqn. \ref{eq:rpath}, yet if $\vperp'$ is too large its gyroradius becomes too large for trapping to occur. Similarly, $\vpar'$ is constrained by the necessity for reflected particles to be able to outrun the shock. The pitch angle, $\alpha' = \arctan(\vperp'/\vpar')$, is a useful quantity that can provide important information about the sub-sample of reflected electrons.

\begin{figure}
    \centering
    \includegraphics[width=\columnwidth]{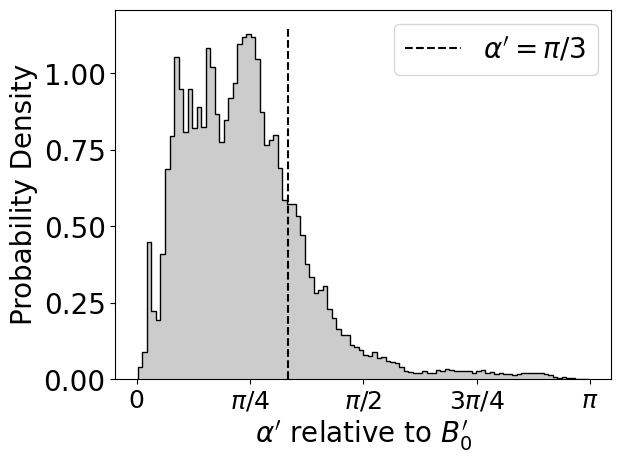}
    \caption{Pitch angle distribution for the most energetic 70 traced particles which have $\gamma \geq 5$ over the final $2\omci^{-1}$ of the simulation. In this frame, all electrons with pitch angle $\alpha \gtrsim \pi/3$ will be caught up by, or are heading towards, the shock.  }
    \label{fig:alpDist}
\end{figure}

We select all \pjm{reflected} electrons that reach $\gamma \geq 5$ at the end of the simulation at $\omci t = 51.5$ and compute their pitch angles as measured in the upstream reference frame relative to $B_0'$ over the final two $\omci^{-1}$ in the simulation. We show the probability distribution of these pitch angles in Fig. \ref{fig:alpDist}. By definition, electrons with pitch angles $\alpha > \pi/2$ are heading back the shock, however, in this frame recall that $\vpar' \cos \thbn' > \vsh$ is required for electrons to outrun the shock and travel upstream. This, in addition to the speed limit of $c$, sets $v^{\prime}_{\parallel, \rm{min}} = 0.53 c$ and $v^{\prime}_{\bot,\rm{max}} \approx 0.85 c$, from which we can calculate a maximum permitted pitch angle of $\alpha_{\rm max}' = \arctan (v^{\prime}_{\bot,\rm{max}}/v^{\prime}_{\parallel, \rm{min}})$ for reflected electrons that are traveling towards the upstream in the upstream rest frame. Accordingly, all electrons with pitch angles $\alpha' \gtrsim \pi/3$ will be caught up to by the shock.

Integrating the probability density shown in Fig. \ref{fig:alpDist} gives the probability of $\alpha' > \pi/3$ as $P(\alpha' > \pi/3) = 0.20$, where this probability represents an estimate of the fraction of reflected electrons that are re-directed back to the shock as a consequence of interactions with the non-linear structures. Considering the stable late-time electron reflection rate of $\approx 4\%$, we estimate that around $0.8\%$ of all the incoming upstream electrons can be energized by the enhanced SSDA mechanism discussed here. \pjm{This argument assumes that the reflected electrons considered here are located within the upstream region containing the non-linear structures, which is true for the majority of reflected electrons within this sample, but the numbers presented here should nevertheless be considered as an approximate upper limit.} 


\subsection{The electron downstream distribution}

\begin{figure}
    \centering
    \includegraphics[width=\columnwidth]{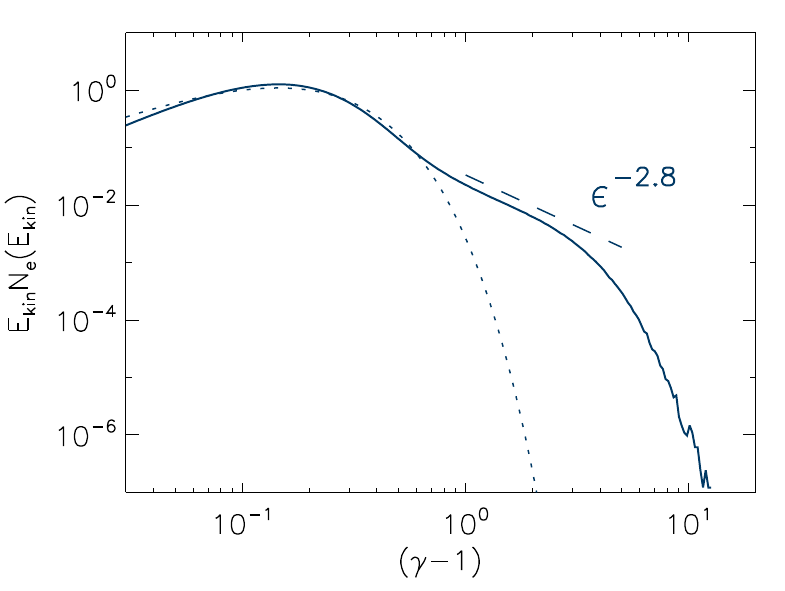}
    \caption{The electron energy distribution in the shock downstream. The dotted line is the Maxwellian fit to the low-energy part of the electron distribution. The dashed line represents the slope of the nonthermal electron tail.}
    \label{fig:spectra-obl}
\end{figure}

Figure~\ref{fig:spectra-obl} shows the electron downstream distribution. The low energy part of the spectra is represented by a Maxwellian  distribution. The electron to ion temperature ratio is about $\te/\ti = 0.17$ which is close to that in a perpendicular shock simulation with similar shock parameters \citep{Bohdan2020b}. 
High energy electrons follow a power-law with an approximate index up to -2.8. The fraction of electrons not covered by the thermal distribution is about $0.8\%$ and they hold about $7\%$ of the downstream electron energy. \pjm{As shown in Fig. \ref{fig:spectra-obl}, at the end of the simulation,} the most energetic \pjm{downstream} electrons \pjm{present at the very tail of the cut-off} already reach gyroradii that are comparable to the shock width or the ion upstream gyroradius, $\gamma_{\mathrm{inj},e} \approx \displaystyle \frac{\mi}{\me} \frac{\vsh}{c} \approx 13$. The threshold for injection into DSA will be a factor of a few higher than that. Unfortunately, the simulation time is still too short to properly capture the DSA phase of acceleration both for electrons and ions, as well as formation of a nonthermal tail in the ion distribution.

\section{Conclusions}\label{sec:conc}

In this paper, we have presented results on the electron foreshock of an oblique collision-less shock, based on a large scale PIC simulation with a total run-time characterized by $\omci t = 51.5$. The physical properties of this simulation are such that a shock front is generated in the high Mach number regime, and thus consistent with the environments of supernova remnants. We have discussed the onset and characteristic properties of whistler waves, and outlined their development into complex non-linear structures which carve out density cavities in the plasma. These non-linear structures are capable of trapping and scattering reflected electrons, confining them to the region close to the shock where they can be efficiently pre-accelerated, which increases the likelihood of being injected into DSA and being further energized. Our main conclusions can be summarized as follows:

\begin{itemize}
    \item Reflected electrons at the shock excite the oblique whistler instability, generating electromagnetic whistlers waves in the inner foreshock region. 
    \item The phase and group velocities of the whistlers are much less than that of the upstream bulk speed $v_{\rm up}$, so they approximately co-move with the upstream plasma and also grow in size to about five ion inertial lengths as they approach the shock. 
    \item Over time, the internal structures arising from the whistler waves become highly complex and non-linear, and they carve out density cavities. The magnetic energy density is the highest at the periphery of each structure, and they are filled with strong electric field that varies on small scales on the order of the ion inertial length. 
    \item Upstream electrons co-moving with the non-linear structures in general experience a Lorentz force directed away from them. This means that upstream electrons in general do not resonate with the structures, thus are unlikely to be confined within them. 
    \item For reflected electrons encountering the non-linear structures, the Lorentz force is now directed towards the structure center, so they can become trapped if their gyroradii are small enough. This enables, and indeed leads to, the return of a sub-set of electrons that are initially reflected at the shock being trapped and returned to it, where additional pre-acceleration is possible.
    \item The acceleration mechanism is directly analogous to stochastic shock drift acceleration, but is around 3 times more efficient. This is because other components of the electric field, besides the coherent motional electric field, can contribute to the acceleration.  
    \item By considering their pitch angles, at any given time, we estimate that $20\%$ of reflected electrons have been redirected back towards the shock. This corresponds to around $0.8\%$ of the total upstream electrons.

\end{itemize}

\acknowledgments

M.P. acknowledges support by DFG through grant PO 1508/10-1. A.B. and M.P. thank the International Space Science Institute (ISSI) for their hospitality and the ISSI Team \textit{Energy Partition across collisionless shocks} for invaluable discussions. The numerical simulations were conducted on resources provided by the North-German Supercomputing Alliance (HLRN) under project bbp00033.

\bibliographystyle{apj}
\bibliography{Bibliography}

\end{document}